# Mapping domain junctions using 4D-STEM: toward controlled properties of epitaxially grown transition metal dichalcogenide monolayers

Djordje Dosenovic[1], Samuel Dechamps[1], Celine Vergnaud[2], Sergej Pasko[3], Simonas Krotkus[3], Michael Heuken[3], Luigi Genovese[1], Jean-Luc Rouviere[1], Martien den Hertog[4], Lucie Le Van-Jodin[5], Matthieu Jamet[2], Alain Marty[2] and Hanako Okuno[1*]

[1] Univ. Grenoble Alpes, CEA, IRIG-MEM, 38000 Grenoble, France
[2] Univ. Grenoble Alpes, CEA, CNRS, Grenoble INP, IRIG-Spintec, 38000 Grenoble, France
[3] AIXTRON SE, Dornkaulstraße 2, 52134 Herzogenrath, Germany
[4] Univ. Grenoble Alpes, Institut Néel, CNRS, 38000 Grenoble, France
[5] Univ. Grenoble Alpes, CEA, LETI, 38000 Grenoble, France

*E-mail: hanako.okuno@cea.fr

**Abstract**

Epitaxial growth has become a promising route to achieve highly crystalline continuous two-dimensional layers. However, high-quality layer production with expected electrical properties is still challenging due to the defects induced by the coalescence between imperfectly aligned domains. In order to control their intrinsic properties at the device scale, the synthesized materials should be described as a patchwork of coalesced domains. Here, we report multi-scale and multi-structural analysis on highly oriented epitaxial $WS_2$ and $WSe_2$ monolayers using scanning transmission electron microscopy (STEM) techniques. Characteristic domain junctions are first identified and classified based on the detailed atomic structure analysis using aberration corrected STEM imaging. Mapping orientation, polar direction and phase at the micrometer scale using four-dimensional STEM enabled to access the density and the distribution of the specific domain junctions. Our results validate a readily applicable process for the study of highly oriented epitaxial transition metal dichalcogenides, providing an overview of synthesized materials from large scale down to atomic scale with multiple structural information.

**1. Introduction**

Growing highly crystalline two-dimensional (2D) transition metal dichalcogenides (TMDs) has become one of the most serious challenges for the realization of ultrathin tunable device applications [1–4]. However, synthesized materials contain generally inevitable intrinsic atomic defects, which are the cause of a discrepancy in the properties between those measured in synthesized materials and the ones theoretically predicted from a perfect model system. Epitaxial growth on a single crystal substrate has become one of the most promising techniques to grow highly crystalline large-scale 2D materials [5]. Coalescence of domains independently nucleated with a single orientation, guided by the single crystalline epitaxial substrate, might result in highly oriented large-scale continuous films. In recent years, a significant progress has been demonstrated in epitaxial growth of 2D materials. In particular, large efforts have been made to reduce the orientation distribution and the amount of the antiphase domains by controlling the substrate surface structure, resulting in nearly single orientation monolayers [6–13]. However, a perfect control of nucleation is difficult due to the weak interaction between the substrate and grown material. As a result, misorientation, misalignment and polar inversion induce domain junctions interfering the seamless coalescence. Successful unidirectional nucleation is often examined by the observation of the orientation of isolated triangular domains at early stage of the growth [10,12] but no follow-up analysis methods have been established for monitoring the density and the distribution of residual defects formed at coalesced domain junctions in continuous layers. Aberration corrected scanning transmission electron microscopy (AC-STEM) has been used as the most powerful technique to characterize atomic defect structures in atomically thin 2D layers. Numerous previous works have revealed possible defect structures such as vacancies and grain boundaries, appearing in synthesized TMDs, associating them with theoretical energy calculations by density functional theory (DFT) [14,15]. However, the fundamental defect study is limited to individual case studies at atomistic level. When we aim at large scale functional materials, the materials should be considered as a patchwork of coalesced domains to control final properties, differed from those expected for a perfect single crystal. In addition, the coalescence of domains with slight misorientation and misalignment induces an important perturbation in the crystal network in medium and long range and can consequently have a large impact, for instance, on the control of stacking and twist angle in future van der Waals heterostructures potentially targeted using epitaxial growth [16]. Therefore, identifying the atomic structure of typical domain junctions and collecting information on their distribution in larger scale will be a key to create a large realistic material model to understand the properties of a real device material. For this purpose, an efficient analytical process providing a direct link between local atomic structure and micron scale order distribution should be established. Selected area-dark field (SA-DF) imaging in conventional TEM mode is commonly used to visualize specific crystal information in relatively large area. The targeted structural features, such as crystal orientation and/or phase can be illuminated in micron scale, where the corresponding diffraction spots are selected using an objective aperture mechanically inserted in the back focal plane [17,18]. The separation of individual electron diffraction spots is not always possible, especially for small angular shift appearing in highly oriented epitaxial layers. Recording the diffraction pattern in probe scanning mode (STEM), local structural information is independently collected at each scanning position, so-called four-dimensional (4D)-STEM acquisition [19]. Analyzing appropriate electron diffraction spots in diffraction patterns enables to map various local structural information, such as orientation, phase and strain [20,21]. The recent evolution in electron sensitive pixelated detector allows quantitative measurement of diffraction signals arising from atomically thin 2D layers [22,23].

In this work, we performed multi-scale and multi-structural analysis of epitaxial TMD layers. Two different materials are studied: $WS_2$ monolayers grown on wafer scale c-plane sapphire by metal-organic chemical vapor deposition (MOCVD) [24] and $WSe_2$ grown on cm scale mica substrate by molecular beam epitaxy (MBE) [25]. The atomic scale analysis on these materials identified typical defect structures. Among the defects present in the materials, we focused on domain junctions directly related to the nucleation and the resulting coalescence. All identified domain junctions could be classified into two types: rotational and polar inverted junctions. The density and distribution of these domain junctions can be considered as the knob parameters for tuning material properties. Multi-structural mapping was realized on wafer scale $WS_2$ monolayer in order to localize and quantitatively analyze the rotational and the inversion domain junctions. These results were correlated to the angular distribution estimated by the Grazing Incidence X-ray diffraction (GI-XRD) [26]. This demonstrated a new way to provide structural overview of epitaxial layers and information on their crystal quality, complementary to existing analytical methods. The multi-structural mapping technique was also applied to visualize orientation and phase in polymorphs (1H and 1T') in MBE grown $WSe_2$ in order to demonstrate potential use of the structural mapping for extended applications.

## 2. Methods

Continuous $WS_2$ monolayer was synthesized onto a wafer scale c-plane sapphire substrate (2 inch) in an industrial MOCVD chamber [24]. $WSe_2$ was synthesized by MBE onto a mica substrate (1.5 x 1.5 $cm^2$), where separated domains consist of monolayer and partially bilayer regions. Both detailed growth process have been previously reported [25]. A $WSe_2$ layer with the angular distribution ~7° = full width at half maximum (FWHM) and $WS_2$ layer with the angular distribution ~4° = FWHM, measured by GI-XRD, are selected for this study in order to find out and validate the structural characteristics in both $WSe_2$ and $WS_2$ epitaxial layers with small in-plane angular distributions. The $WS_2$ and $WSe_2$ systems were transferred using a polymer support onto TEM grids by water capillary delamination. For the observation of non-continuous $WSe_2$, a commercial CVD graphene (Graphenea) was pre-transferred on the TEM grid as an electron transparent support layer. The transferred TMD layers are then analyzed by high angle annular dark field (HAADF)-STEM using Titan Ultimate and Titan Themis Thermofisher working at 80 kV and 200 kV. Structural maps are reconstructed from diffraction patterns acquired at each probe position using a direct electron detector (Quantum

## 3. Results and discussions

### 3.1 Identification of epitaxial domain junctions

First, detailed atomic resolution analysis was performed on both MOCVD grown $WS_2$ and MBE grown $WSe_2$ samples in order to identify the defect structures specific to the epitaxial growth. Several defects such as vacancies or holes are observed, some of them can be removed by thermal annealing under sulfur or selenium atmosphere [27]. Here we focused on identifying the typical domain junctions formed during epitaxial growth. Figure 1 shows two typical domain junctions appearing in highly oriented epitaxial $WSe_2$ layers. The atomic resolution image demonstrates the presence of domains with different polar directions indicated as A and B (figure 1(a) and 1(b)). The geometrical phase analysis (GPA) based on Fast Fourier transform (FFT) was then used to map local orientations in the image (figure 1(c)). Here, three rotational junctions are revealed including the inversion domain boundaries (IDBs) already identified in the atomic resolution image. Interestingly, an in-plane crystal rotation of 3° was detected in a structurally continuous domain. This rotational junction is thought to be formed by a coalescence of misoriented neighboring domains. A quasi-perfect coalescence between two rotated domains was also confirmed in atomic resolution analysis, as shown in figure 2(a), up to 5° of misorientation in $WSe_2$ monolayer. Here the 5° misorientation is accommodated by local strain and dislocations forming at the domain junction. The Fourier filtered image shows the missing half-plane and the dislocation cores (in the inset). This type of junction is classified as low-angle (LA) GBs in the literatures [28]. On the contrary, two inverted domains (A and B), are connected with chain boundaries (figure 2(b)), where their crystal structure is totally interrupted by a 1D line defect. The IDBs observed in $WSe_2$ monolayer represent the type 4|4P and 4x|8 line structures, whereas the number of 4 rings between two 8 rings varied as already observed in $MoS_2$, $WSe_2$ and $WS_2$ [15,29,30].

The same results were obtained in $WS_2$ monolayers. Our structural screening revealed that misoriented and non-inverted domains coalesce with a small amount of dislocation cores and local strain but without any line defects as shown in figure S1. It should be noted that the misorientation of around 2° and misalignment between two areas are sometimes simply compensated by the presence of holes or lattice deformation. This observation gives a conclusion that misorientation and misalignment between non-inverted domains cause only a few dislocation cores, holes and local strain around the coalesced junctions. IDBs in $WS_2$ also showed the same structural characteristics as those clearly observed in $WSe_2$ by Z-contrast images. Figure 3 shows typical atomic image of boundaries between two inverted domains identified in $WS_2$. The interpretation of the Z-contrast image of $WS_2$ requires a particular attention. Different from the WSe$_2$ monolayer with heavy chalcogen atoms giving significant contrast, the determination of polar direction by atomic resolution images in WS$_2$ consisting of heavy tungsten and light sulfur atoms is sensitively impacted by three-fold astigmatism. In spite of improved aberration correction technologies, the residual three-fold astigmatism often exceeds the threshold to artificially inverse local polarity of one side of inverted domains in a Z-contrast image with large convergence angle, leading to a wrong interpretation (figure S2). Therefore, a relatively low convergence angle (semi-angle = 18 mrad) was used for structural determination of IDB structures with minimized aberration effects. The detailed aberration effect on the image contrast in WS$_2$ is described in the Supplementary Information. The atomically resolved HAADF image clearly confirmed the phase inversion on the two sides of a domain boundary. The inverted domain junctions found in WS$_2$ monolayers consist of line defects with a combination of 4 and 8 rings in all observed cases as the example shown in figure 2(b) for WSe$_2$. Recently a new type of non-inverted grain boundary appearing in epitaxial WS$_2$ monolayer has been reported [31]. Our experimental results confirm that all GBs observed in this work are the IDBs with 4x|8 types of line defects. The number of continuously connected 4 rings between 8 rings varies from one up to more than 10 as shown in figure 3(a)-(c), depending on the geometry of GBs. The complementary DFT calculations showed that the most stable inversion domain boundary in WS$_2$ is the one without any 8 rings, so-called 4|4P line defect [30]. However, the 4|4P is a perfect mirror twin boundary (MTB), therefore a perfect local alignment and a straight mirror plane are required to form this stable structure. Contrary to the MTBs often observed in MoSe$_2$, which form triangular patterns inside single crystals due to the Se deficiency in the growth process [32], the topology of IDBs in WS$_2$ might result from the competition between the misalignment, the misorientation and the front geometry of the coalescing polar inverted domains. The formation energies of 4x|8 IDBs (figure 3(d)) are calculated using DFT in the W-rich (S-deficient) and S-rich (W-deficient) limits. Computational details are given in the Supplementary Information. Figure 3(e) depicts the energetics of 4|4P line defects containing 8 rings, providing substantial insight into the role of those non-hexagonal rings on stability. First, we report a value of 0.98 eV for the 4|4P formation energy in the W-rich phase, in excellent agreement with previous theoretical work [33]. Although arrays of only 8 rings are significantly more costly, the formation energy of line defect decreases with increasing number of 4 rings added between 8 rings. Finally, pairs of 8 rings separated by more than two 4 rings become energetically comparable to the stable 4|4P. These results confirm that the 4x|8 lines are formed at all inverted domain junctions because of thier energetic stability and geometrical flexibility, allowing to connect misaligned, misoriented and curved domain junctions. After the screening of atomic defects, we conclude that typical domain junctions in highly oriented epitaxial WSe$_2$ and WS$_2$ can be classified into two types. i) the rotational domain junctions with very few dislocations and local strain but without any linear defect breaking the crystal continuity (this is confirmed for up to 5° of misorientation) and ii) the IDBs consisting of 4x|8 based line defects.

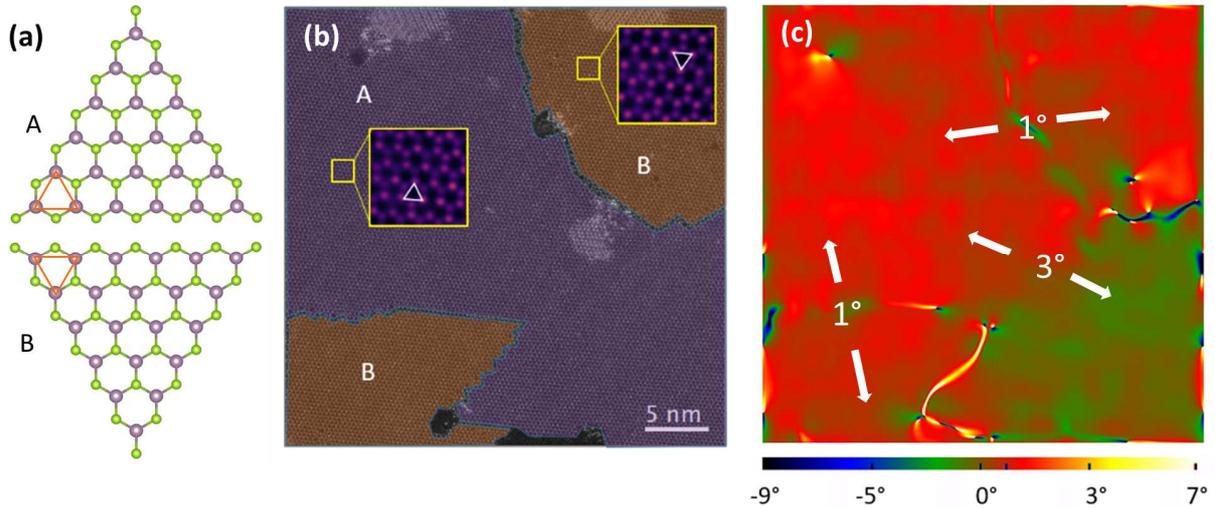

*Figure 1: Typical domain junctions appearing in highly oriented epitaxial layers. (a) Atomic model of two domains with opposite polar direction denoted as A and B (b) Atomic resolution HAADF STEM of $WSe_2$ monolayer grown on mica substrate, indicating the presence of two polar inverted domains A and B and (c) orientation map constructed by GPA indicates the presence of rotational domain junctions.*

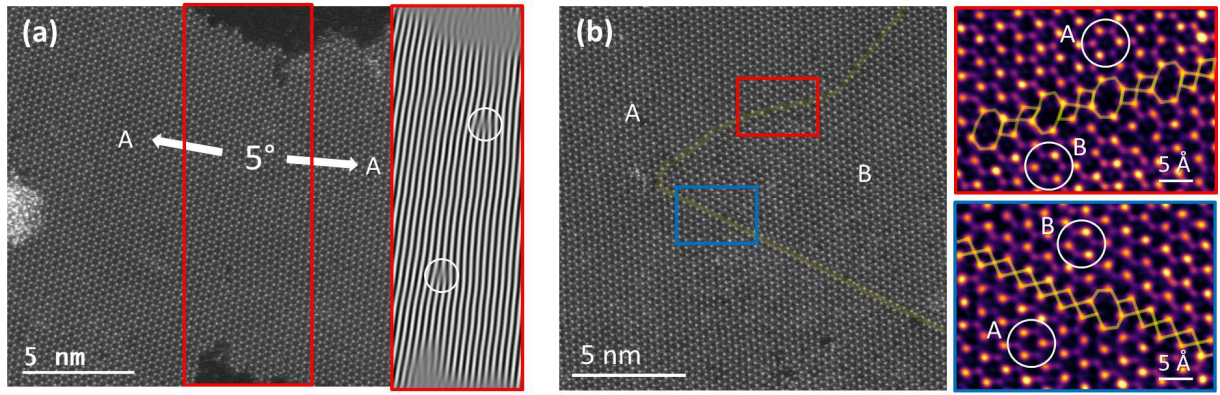

*Figure 2: HAADF images of atomic structures of WSe$_2$. (a) 5° rotated domain junctions. Two non-inverted domains are coalesced with 2 dislocation cores highlighted by white circles in the Fourier filtered image (inset) and (b) boundary between two domains with opposite polar directions. The insets show low-pass filtered images of (upper) curved GB with 2° rotation and (bottom) straight GB with 0° rotation.*

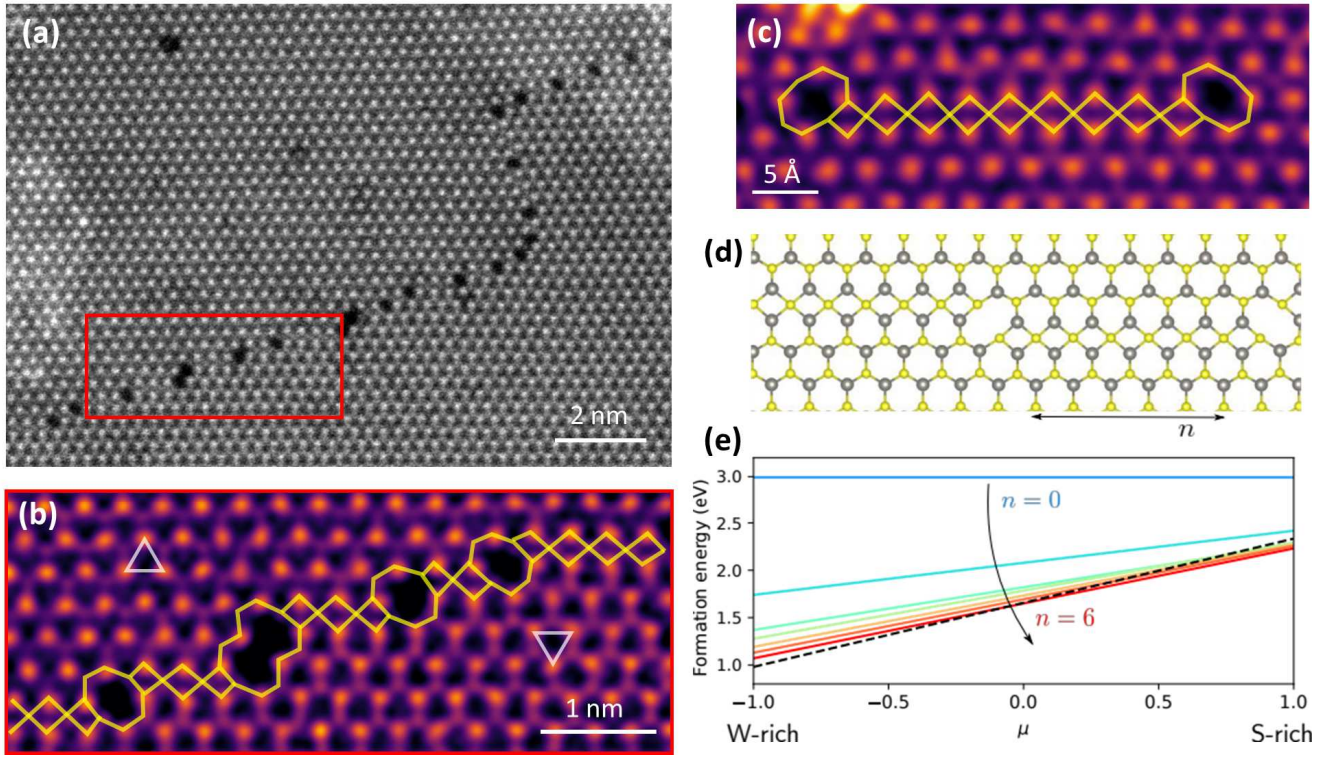

*Figure 3: (a-c) HAADF images of polar inverted junction in MOCVD grown WS$_2$ continuous layer. Atomic resolution image shows clearly the inversion of polar direction between two domains separated by 4x|8 type line defects. (d) DFT structural model of 4x|8 IDB and (e) DFT calculated formation energies for different number n of 4 rings between two 8 rings. The black dashed line concerns the formation energy of the reference 4/4P mirror twin boundary.*

*3.2 Visualization and quantification of domain junctions*

The electronic properties of GBs strongly depend on the misorientation angle [34]. Low angle GBs with individual dislocation cores, as observed in epitaxial WSe$_2$ and WS$_2$, are expected to have little impact on the electronic properties. In this configuration, carriers simply pass within coalesced crystals, leading to improved mobility by reducing misorientation angle and the density of misoriented domain junctions [35]. On the contrary, the density of IDBs, 1D metallic strips [36], might strongly influence the transport properties of layers in large scale. In the following, we visualize the spatial distribution of these domains using 4D-STEM structural mapping method.

*3.2.1 Orientation mapping*

Figure 4 shows an analytical process carried out on WS$_2$ monolayer grown on wafer-scale c-plane sapphire by MOCVD. In general, GI-XRD performed directly on as-grown samples is a powerful tool to determine the in-plane angular distribution together with the epitaxial relationship compared to the underlying substrate. Here, GI-XRD results on an example of as grown WS$_2$ is presented in figure 4(d). The azimuthal scans confirm the highly anisotropic angular distribution. The FWHM of the peaks, which reflects the in-plane misorientation of the grains, amounts to $\Delta\Phi = 3.85\pm0.2°$ by averaging over (020) and (110) peaks. GI-XRD shows the good epitaxial growth with alignment between sapphire <10$\bar{1}$0> and WS$_2$ <11$\bar{2}$0> axes with an angular distribution <4° (FWHM). The same sample was transferred onto TEM grid (figure 4(b)) and analyzed by 4D-STEM. A series of diffraction patterns was acquired for all positions of the probe scanning the sample surface. Figure 4(e) shows the diffraction pattern averaged over all real space pixels. Each recorded diffraction image is then labelled with the pixel position and the corresponding orientation angle using 6-fold symmetry information in the diffraction pattern. It provides a real-space orientation map (figure 4(c)) and a histogram of angular distribution (figure 4(f)) over the analyzed area. The detailed mapping method is described in the following section. This process allows immediate comparison to the XRD data to be able to: i) find an area of interest that represents the mm scale fingerprint of the sample and/or ii) verify the uniformity of the samples by using different samplings on the wafer. Here the angular distribution (FWHM=3.3°) measured in a ~4 µm$^2$ area is in good agreement with XRD data. In the orientation map in figure 4(c), parallel lines are observed that correspond to repetitive small angle rotation spaced with an interval of 45 nm. The distance corresponds to the terrace width for monoatomic step edges of sapphire with a miscut angle of 0.27° determined using XRD. According to XRD data, the direction of these lines (WS$_2$ <10$\bar{1}$0>) is perpendicular to the sapphire <10$\bar{1}$0> direction that can correspond to the miscut direction of the sapphire substrate. The repetitive rotation of WS$_2$ domains coincides with the typical length of the substrate terraces as well as with their direction established by XRD. The influence of sapphire step edges on the local domain rotation is thus visualized by the orientation mapping.

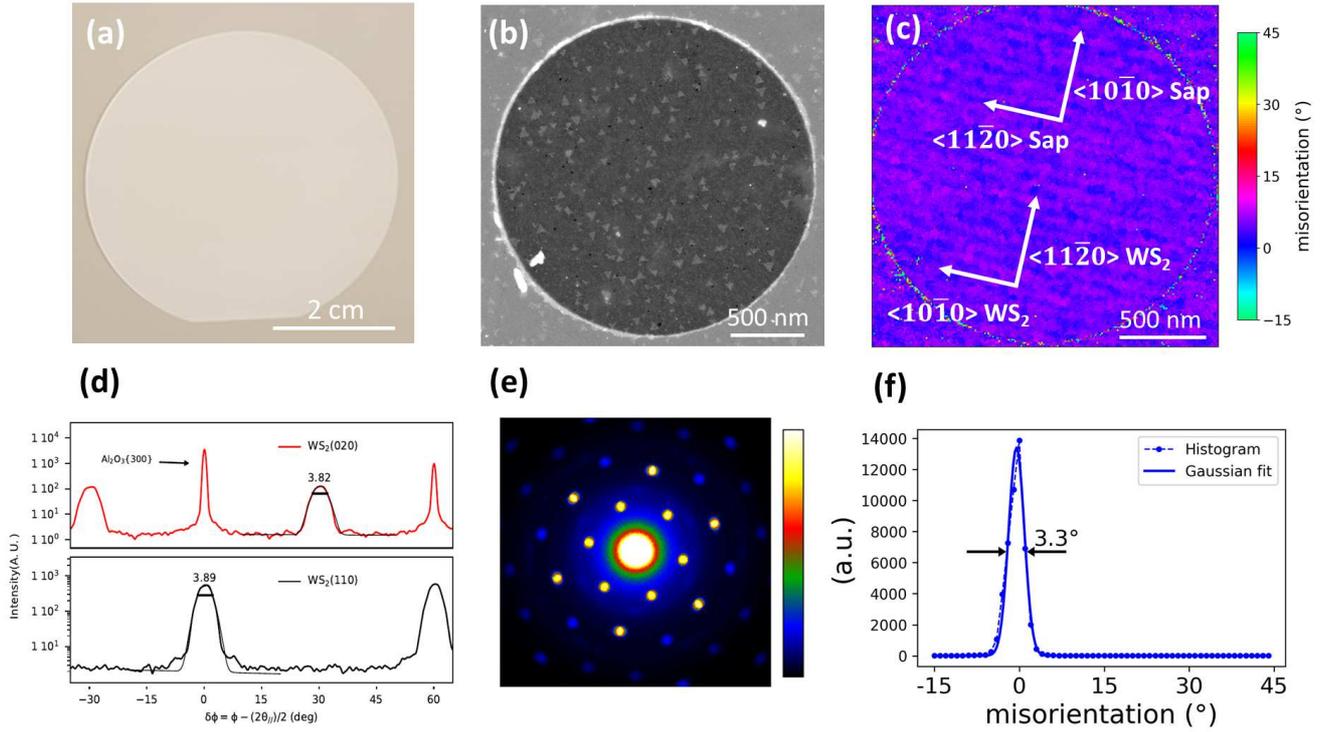

*Figure 4: Orientation study of WS$_2$ monolayer grown on c-plane sapphire by MOCVD. (a) As grown WS$_2$ layer on wafer scale (2inch) sapphire substrate. (b) HAADF image of free-standing WS$_2$ monolayer transferred onto TEM grid. (c) Orientation map reconstructed from diffraction dataset acquired by 4D-STEM acquisition. (d) X-ray diffraction measured on as grown sample. (e) Diffraction pattern averaged over all real space pixels. (e) Histogram of orientation measured in each diffraction pattern (48400 real space pixels).*

*3.2.2 Polar asymmetry mapping*

The orientation map in figure 4 reflects only hexagonal direction without distinction in polar directions and is analogous to the orientation distribution provided by XRD. Recently, a possible determination of polar asymmetry has been demonstrated by the anomalous contrast between $hkl$ and $\overline{hkl}$ peaks in diffraction pattern of TMD monolayers. These materials act as strong phase object even in monolayer limit and break Friedel's law that generally imposes in-plane inversion symmetry of diffraction pattern for thin and light materials within weak phase object approximation [37]. Especially, the identification of polar direction in electron diffraction has been shown to be quite reliable for 2D materials with pronounced polarity i.e. containing elements with a large difference in atomic number, such as WS$_2$. In order to localize inversion domain junctions, we labelled orientation angles using 3-fold symmetry in the diffraction dataset. Figure 5 shows the multi-structural mapping including orientation and polar information, performed on two WS$_2$ samples with the same angular distribution but with different structural morphologies. As shown in figure 5(a) and 5(b), we detected the three intense spots in diffraction pattern indicating polar direction. To generate the multi-structural maps, the orientation and polar information was extracted by a custom-made template matching algorithm based on the work in ref [38]. A template diffraction pattern $T(h,k)$ is chosen directly from the acquired 4D dataset. The template diffraction pattern is then reduced to the pixels $h, k$ containing only the 1$^{st}$ and 3$^{rd}$ order reciprocal spots where the inversion symmetry is broken. Such a reduced template is then rotated in the range [0°, 120°) with a step of 1°. A total of 120 template diffraction patterns are created and each recorded diffraction pattern $P(h,k)$ is compared to a created set of 120 templates $T_i(h,k)$ by calculating the correlation index as follows:

$$Q_i = \frac{\sum_{hk} P(h,k)\, T_i(h,k)}{\sqrt{\sum_{hk} P^2(h,k)} \sqrt{\sum_{hk} T_i^2(h,k)}}$$

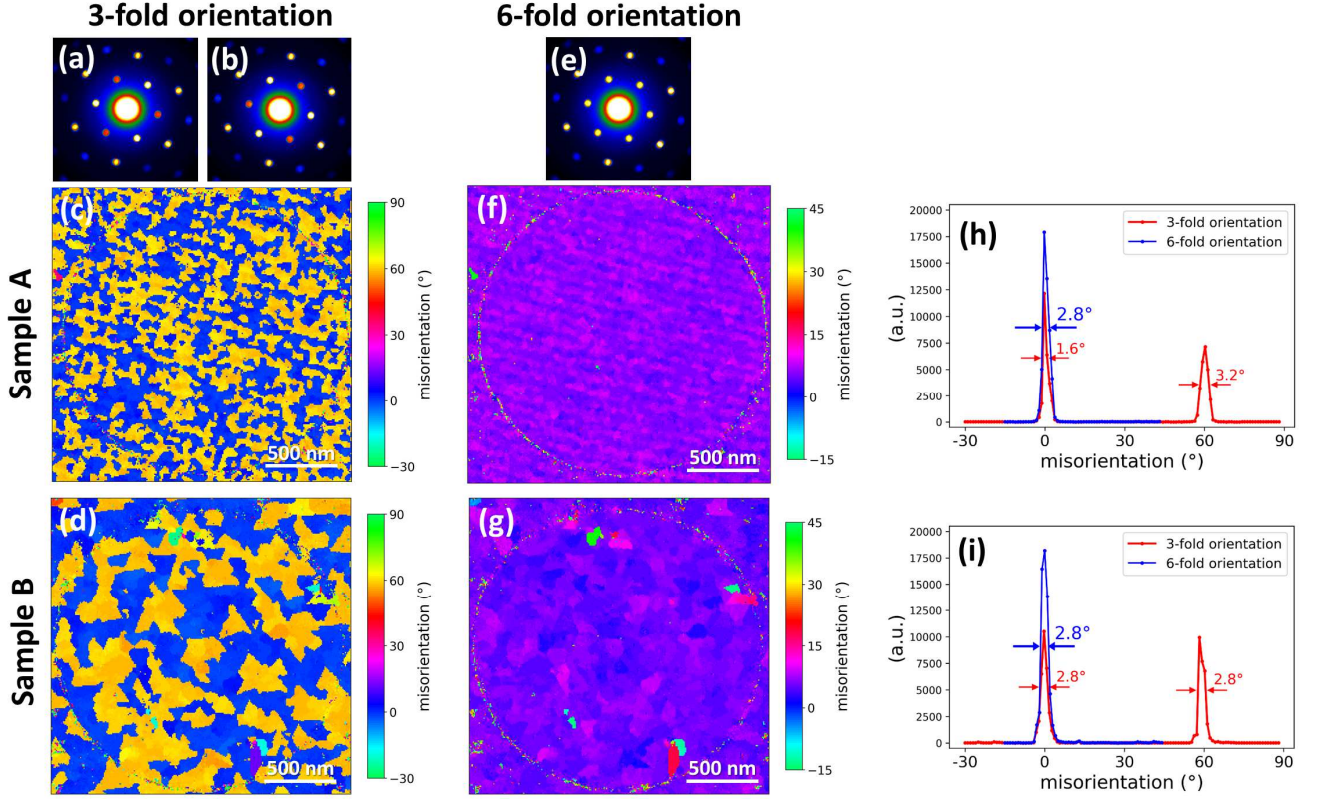

*Figure 5: Multi-structural mapping of WS$_2$ monolayer grown on c-plane sapphire by MOCVD. (a-b) 3-fold rotation in diffraction pattern averaged over only one polar direction used for inversion domain maps of (c) sample A and (d) sample B. (c) 6-fold rotation in diffraction pattern averaged over all real space pixels and corresponding orientation maps of (f) sample A and (g) sample B. Histograms of simple 6-fold rotation corresponding to the orientation distribution (blue) and 3-fold rotation (red) where two peaks correspond to the orientation distribution of each antiphase for two different samples: (h) sample A and (i) sample B.*

The orientation of the 3-fold template yielding the highest correlation index is attributed to the local crystal orientation including polarity distinction at the given beam position (-30°<= θ <90). This provides orientation maps including information on polar direction as shown in figure 5(c) and 5(d). The 6-fold orientation maps can be subsequently obtained by modulo 60° of the 3-fold information (θ mod 60°), shown in figure 5(f) and 5(g), directly comparable with XRD data. According to the histograms of 6-fold orientation (blue curves in figure 5(h) and 5(i)), both samples exhibit comparable angular distribution with a FWHM=2.8°. This information is already accessible by the XRD analysis while the 4D-STEM mapping additionally provides spatially resolved angular distribution. Moreover, the histograms of the 3-fold orientation (red curves in figure 5(h) and 5(i)) reveal the presence of different polarities with the independent angular distributions centered around 0° and 60° and visible as blue and orange domains in the 3-fold orientation maps (figure 5(c) and 5(d)). Despite the similarities in the angular distribution, these two samples exhibit a significant difference in layer morphology, in particular domain size and shape, which may provide important clues to the dominant parameters for the resulting material properties. Based on the high resolution study of domain junctions shown in figure 2, the neighboring domains with different polarities are connected with IDBs identified as 1D 4x|8 line defects. From the 3-fold orientation maps, the density of IDBs for the two samples was quantified as 34 µm/µm$^2$ and 19.5 µm/µm$^2$, respectively. The multi-structural maps also allow superposing two types of domain junctions (as shown in figure S5). The results clearly show that the rotational and polar inverted junctions are not always identical. This spatially and quantitatively resolved information enables to assess growth process ultimately targeting unidirectional nucleation to achieve seamless domain coalescence.

*3.2.3 Phase mapping*

Finally, the multi-structural mapping technique is applied to characterize an epitaxially grown heterophase WSe$_2$:1H-1T' monolayer. The two phases have significantly different electronic properties: WSe$_2$:1H is a semiconductor while WSe$_2$:1T' is a Quantum Spin Hall (QSH) insulator [39,40] and the topologically protected edge states are known to be formed at the interface between the two phases [41]. The properties of such heterophase system therefore depend on the ratio of the different phases as well as on the relative distribution of these phases and junctions between them [42,43]. Figure 6(a) shows a HAADF image of WSe$_2$ grown on a mica substrate by MBE under specific experimental conditions. The presence of metastable WSe$_2$:1T' has been firstly evidenced in-situ during the growth by reflection high energy electron diffraction (RHEED) and further by the atomic scale HAADF imaging (figure 6(f)). As the domains are not perfectly coalesced, the sample was transferred onto graphene supported TEM grid. The phase mapping of the WSe$_2$:1H-1T' sample was done by a template matching algorithm as explained before. Three template diffraction patterns from the dataset were chosen, corresponding to graphene, WSe$_2$:1H and WSe$_2$:1T' phases.

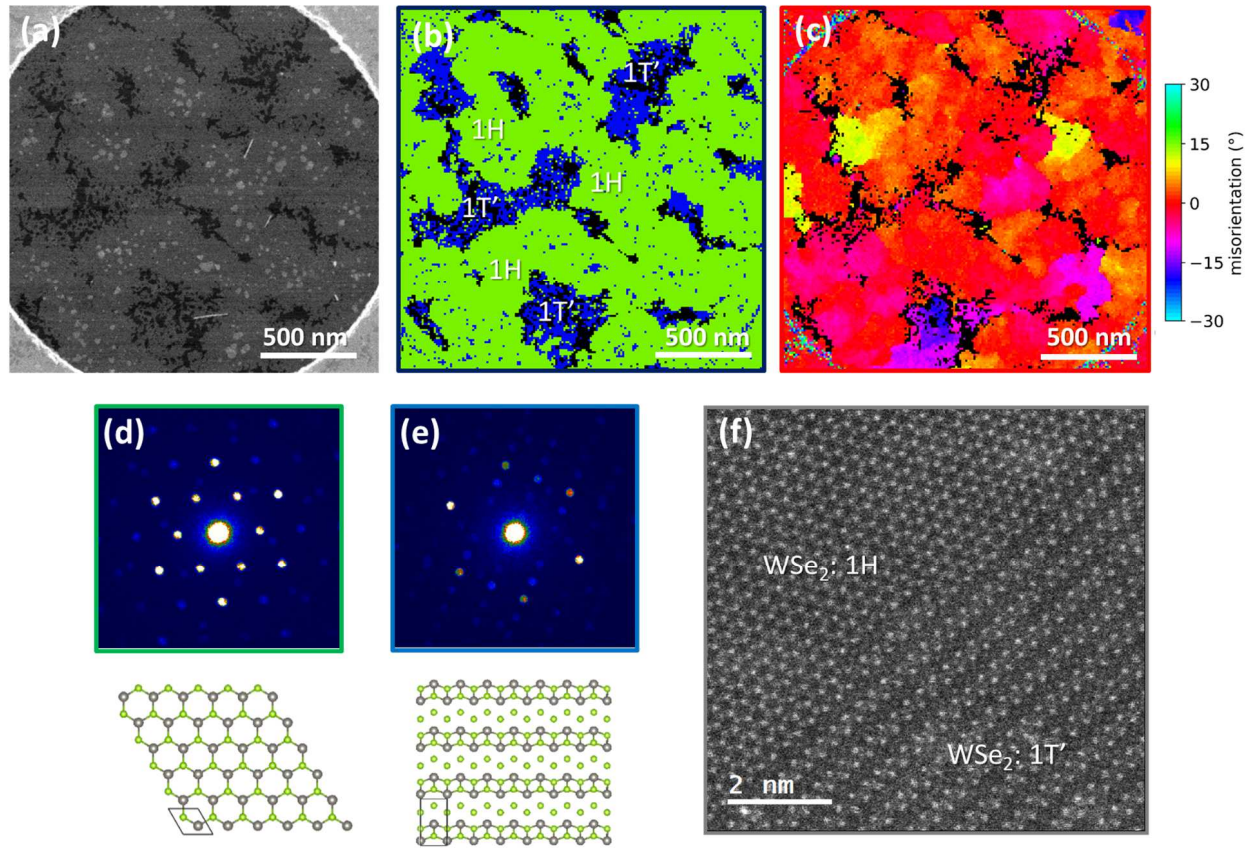

*Figure 6: Multi-structural mapping of WSe$_2$ monolayer grown on mica by MBE. (a) HAADF image of free-standing WSe$_2$ monolayer transferred onto graphene supported TEM grid. (b) Local phase map of 1H phase indicated in green and 1T' phase indicated in blue and (c) corresponding orientation map. Diffraction patterns and atomic structure models of (d) WSe$_2$: 1H phase and (e) WSe$_2$:1T' phase. The black regions in both phase and orientation maps correspond to the graphene support membrane. (f) Atomic resolution HAADF image of phase junctions between 1H and 1T' WSe$_2$.*

The templates are further reduced to a set of pixels h,k containing the diffraction spots characteristic of the given phase. An additional set of templates is created by rotating the three initial templates by a step of 1° in the range [0°, 60°] for graphene and the 1H phase and [0°, 180°] for the 1T' phase, given their 6-fold and 2-fold point group symmetry, respectively. In this case, the possible polarity of WSe$_2$:1H is neglected because of the small size of inverted domains in this specific material. Each diffraction pattern is thus compared to the total of 300 templates and the highest correlation index is found that corresponds to the best phase and orientation match. The detailed data treatment procedure is described in the Supplementary Information. Figure 6 (b)-(c) show the orientation and phase maps reconstructed from the same diffraction dataset. The phase map of a 4 µm$^2$ area shows a significant proportion of 1T' phase ($\approx$ 16 %) that never stands alone, but is rather distributed on the edges of 1H domains. The orientation map from the same region shows that 1T' phase has the same orientation as neighboring 1H domains. The overall orientation agreement between the two phases indicates that 1T' domains did not nucleate independently but were rather created in a phase transition from 1H domains during the growth. The micrometer scale phase and orientation mapping obtained from a single 4D-STEM acquisition is demonstrated as a route to access the above-mentioned information relevant to the material properties as well as to give insight in growth mechanisms of heterophase 2D systems.

## 4. Conclusions

In summary, we demonstrated a multi-scale analytical process for epitaxially grown highly oriented TMD monolayers. The atomic resolution analysis identified two major structural junctions such as low angle domain junction and inversion domain junction. The former represents a relatively perfect coalescence with the presence of few dislocations and was observed for domains misoriented by up to 5°. The latter forms chain-like 1D continuous defects consisting of 8 rings separated by 4|4 chains. The orientation map was successfully demonstrated using diffraction dataset obtained by 4D-STEM acquisition, which allowed to spatially resolve the orientation distribution measured by XRD. In some cases, we observed the impact of substrate surface morphology on local orientation in grown layer. The local polar direction was analyzed using the intensity anomaly appearing in electron diffraction pattern recorded at each pixel point to visualize the inversion domains formed in a continuous WS$_2$ monolayer. This data can be used to quantify the total length of typical 4x|8 units to further enable theoretical studies on a realistic large-scale structural model including defect information. The feasibility of inversion domain mapping was validated for WS$_2$ monolayers. Mapping of polymorphs, 1H and 1T' phases was also demonstrated in WSe$_2$ sample. The multi-structural mapping allowed to correlate the local orientation and phase junction, which revealed epitaxial phase transition of 1H to 1T' phases from the center toward the edge of domains without any rotations. These characterization process can further include extraction of other information such as strain map and stacking and twist angle maps in vdW heterostructures. Increasing the sample size to be studied and the precision of measurements in both real and reciprocal space, the size of data to be treated will become larger. The optimization and acceleration of the characterization process will thus be a key for the efficiency in the study of epitaxial growth of large-scale 2D materials.

## Acknowledgements

We acknowledge the French National Research Agency through the MAGICVALLEY project (ANR-18-CE24-0007). The LANEF framework (No. ANR-10-LABX-0051) is acknowledged for its support with mutualized infrastructure. This project received funding from the European Research Council under the European Union's H2020 Research and Innovation programme via the e-See project (Grant No. 758385). A CC-BY public copyright licence has been applied by the authors to the present document and will be applied to all subsequent versions up to the Author Accepted Manuscript arising from this submission, in accordance with the grant's open acces conditions. These experiments have been performed at the Nanocharacterisation platform in Minatec, Grenoble.

# Supplementary Information:

# Mapping domain junctions using 4D-STEM: toward controlled properties of epitaxially grown transition metal dichalcogenide monolayers

Djordje Dosenovic[1], Samuel Dechamps[1], Celine Vergnaud[2], Sergej Pasko[3], Simonas Krotkus[3], Michael Heuken[3], Luigi Genovese[1], Jean-Luc Rouviere[1], Martien den Hertog[4], Lucie Le Van-Jodin[5], Matthieu Jamet[2], Alain Marty[2] and Hanako Okuno[1*]

[1] Univ. Grenoble Alpes, CEA, IRIG-MEM, 38000 Grenoble, France
[2] Univ. Grenoble Alpes, CEA, CNRS, Grenoble INP, IRIG-Spintec, 38000 Grenoble, France
[3] AIXTRON SE, Dornkaulstraße 2, 52134 Herzogenrath, Germany
[4] Univ. Grenoble Alpes, Institut Néel, CNRS, 38000 Grenoble, France
[5] Univ. Grenoble Alpes, CEA, LETI, 38000 Grenoble, France

*E-mail: hanako.okuno@cea.fr

1. **Low angle domain junctions in WS$_2$**

Figure S1 shows typical structures around rotational domain junctions found in WS$_2$ monolayers. The GPA was processed on each HAADF image acquired on a 200 x 200 nm$^2$ area to identify rotated areas and to study the related atomic defect structures. All non-inverted but rotated junctions are quasi perfectly coalesced with a small amount of dislocation cores and local strains. No line defect is observed between misoriented and misaligned neighbouring domains among numerous analysis performed over the mm-size TEM grid for several samples. The dislocation cores at the convergent point of additional planes consist of 5-7 defects as shown in figure S1(c) and S1(d). Larger holes were also found in many cases at the same position instead of 5-7 defects.

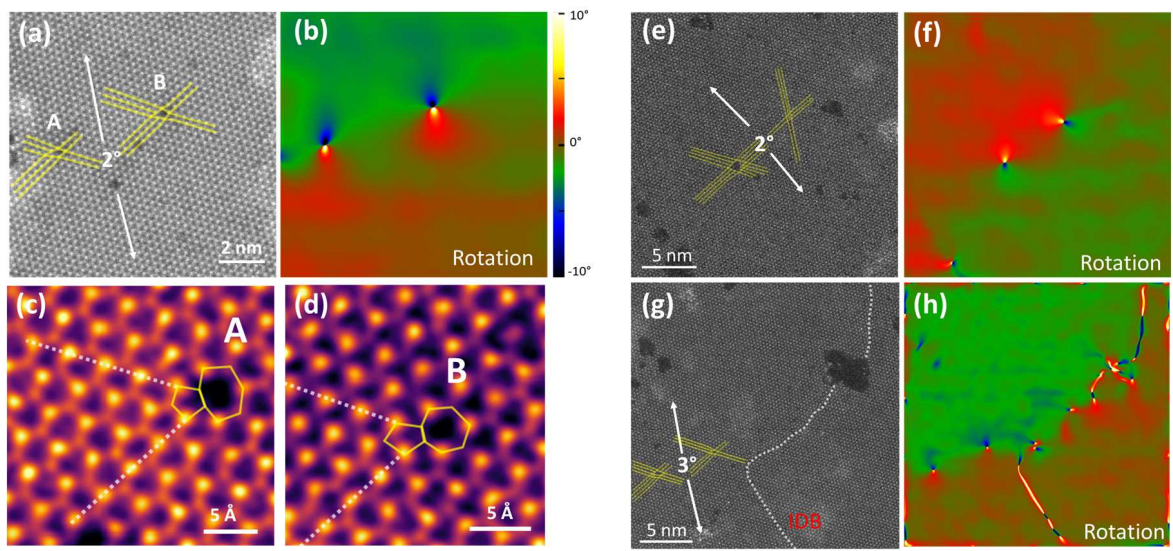

*Figure S1: (a, e, g) HAADF images of rotational domain junctions in the WS$_2$ monolayer (4°= FWHM by XRD), with (b, f, h) corresponding local orientation maps obtained by the geometrical phase analysis (GPA) method. (c-d) atomic images of dislocation cores found in (a).*

2. **Inversion domain boundaries in $WS_2$:**

Figure S2 shows the impact of the residual three-fold astigmatism on Z-contrast images. In general, imaging IDBs with a perfect symmetrical contrast is quite difficult because the residual three-fold astigmatism impacts only one side of IDBs[1,2]. This makes it difficult to determine the local polar direction. In particular, in the case of $WS_2$, the image contrast is sensitively influenced by this aberration, where an artificial bright spot appears at the center of hexagons as shown in the bottom part of the simulated images (figure S2(f)). One way to limit this phenomenon is to reduce the convergence angle down to a value reducing the effect of three-fold astigmatism as shown in figure S2(b).

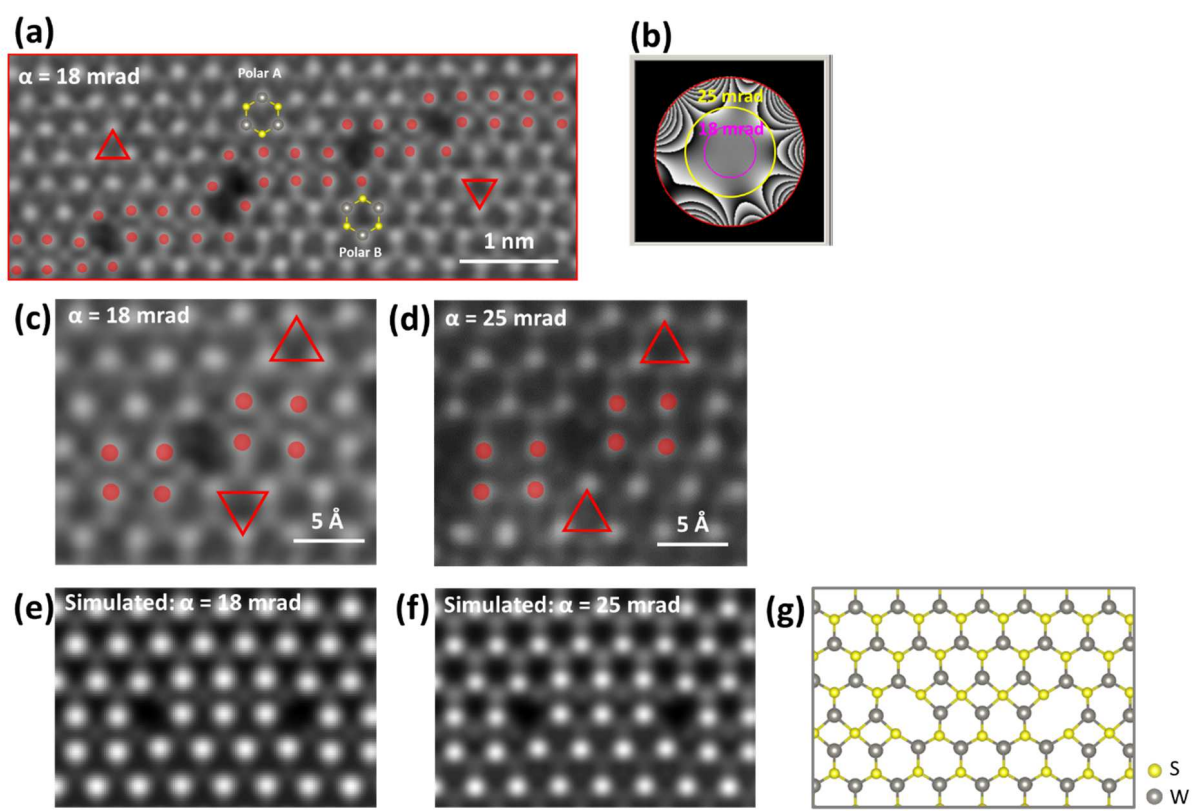

*Figure S2: (a) HAADF image of the typical atomic structure of IDBs. (b) an example of phase plate representing the three-fold astigmatism in the condenser lens compared to the convergence angle. (c-d) Comparison of HAADF Z-contrast images acquired on the same structures with different convergence angles. Red points indicate W atom positions and red triangles indicate local polar directions. (e-f) Comparison of HAADF Z-contrast images with different convergence angles simulated on the same structural model shown in (g).*

### 3. DFT computational information:

First principles calculations are performed within density-functional theory (DFT), norm-conserving pseudopotentials and pseudo-atomic localized basis functions, using the *OpenMX* software package[3–5]. The exchange-correlation type is based on the generalized gradient approximation (GGA), as implemented by Perdew-Burke-Ernzerhof (PBE)[6]. Electron-ion interactions are described using Troullier-Martins-type norm-conserving pseudopotentials[7]. The basis functions are defined for both W and S atoms as *s2p1d1*, meaning that two *s*-, one *p*- and one *d*-orbitals are considered. Applying the Monkhorst-Pack scheme[8], the electronic structure of TMDs are computed with a *k*-points sampling equivalent to a $10 \times 10 \times 1$ mesh in the unit cell. The real-space mesh energy cutoff is set to 300 *Ry*. To avoid spurious interactions, an inter-layer distance of 50 Å is employed between the periodically repeated images. Lastly, an electronic temperature of 300 K was used. The computations yield 3.21 Å for the lattice parameters of $WS_2$, in good agreement with the literature[9,10]. Similarly, a reported band gap of 1.58 eV is consistent with previous DFT calculations[11]. The inversion domain boundaries (IDBs) were structurally optimized until reaching a force criterion of $4 \times 10^{-4}$ Ha/$a_0$ where $a_0$ is the Bohr radius.

### 3.1. Formation energies of IDBs

The formation energy of grain boundaries in the ribbon geometry is defined as

$$E_f = E_{IDB} - (n_W E_W + n_S E_S + 2\gamma)$$

where $E_{IDB}$ is the total energy of the supercell, $n_i$ and $E_i$ are the number of atoms and the chemical potential associated to each species, respectively. The ribbon edges are included through the parameter $\gamma$ that accounts for their formation energy. Importantly, the ribbon geometry offers the advantage to relax the structure perpendicularly to the IDB, in spite of the minimal computational overhead related to $\gamma$. The accessible range of the chemical potentials for W and S atoms are constrained by the equilibrium condition

$$E_{WS_2} = E_W + 2 E_S$$

in which $E_{WS_2}$ is the energy of a primitive cell of $WS_2$. Similarly to Refs.[12,13], the W-rich and S-rich ends are limited by the metal body-centered cubic structure and octasulfur, respectively. Since 4|4P are Se-deficient, we are mostly interested in the W-rich phase. The IDBs observed in $WS_2$ contain various arrangements of 4 and 8 rings. For comparison with 4|4P line defects, the formation energy of any 4x|8 grain boundary is renormalized as

$$E_f^* = E_f/(N_4 + N_8)$$

where $N_4$ and $N_8$ account for the number of 4 and 8 rings in the line defects, respectively.

### 3.2. Edges energies

To estimate $\gamma$ for only one type of edges, the trigonal symmetry of TMDs imposes to employ triangular flakes, as ribbons always contain two different types of edge reconstructions. From its definition, the total energy of the triangular flakes ($E_{tot}$) yields the following relation

$$3N\gamma + 3\beta = E_{tot} - n_W E_{WS_2} - \delta n E_S$$

The flake size ($N$) corresponds to the number of units along one edge, while the corners energy is accounted for by $\beta$. Lastly, $\delta n$ represents the off-stoichiometry from the 1:2 ratio between W and S, bringing dependence on the chemical potential to the edge energy.

Figure S3 depicts the linear fit of the total energy in both limits after computing the ground states of flakes of different sizes. In this case, 100 % chalcogen passivated W edges were considered. Edge energies of 3.57 eV and 0.87 eV are obtained in the W-rich and S-rich ends, respectively, in good agreement with the 2.54 eV and 0.90 eV reported in Ref.[13]

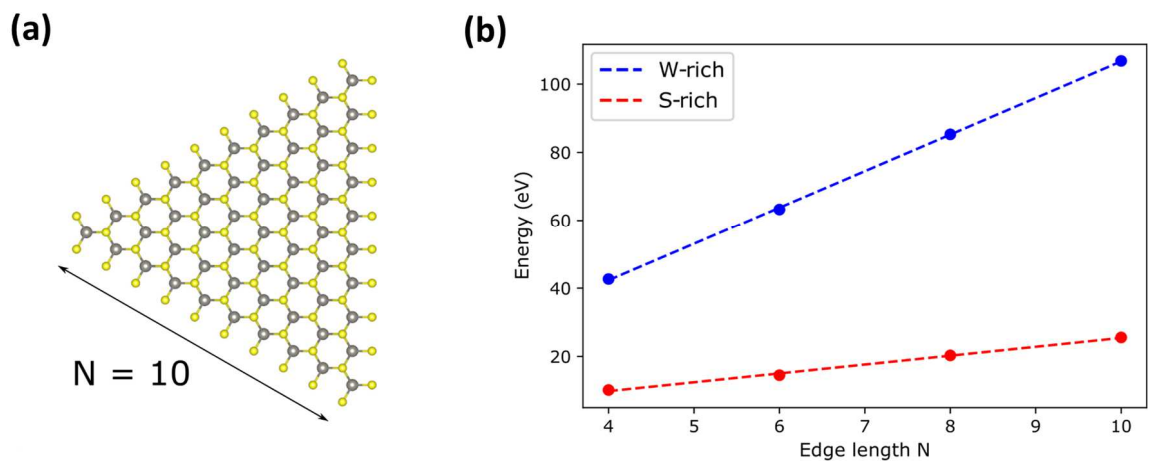

*Figure S3: Edges energies from triangular flakes. (a) Model of a triangular flake with W edges completely passivated with S atoms when N = 10 (b) Energy of the triangular flake as a function of the flake edge length, with the linear fits for comparison (dashed lines). Model visualized using VESTA[14]*

4. **Orientation map of MBE grown WSe₂:**

Figure S4 shows structural characteristics of the WSe$_2$ layer studied in this work. The angular distribution (FWHM) was measured as 7 ° around the epitaxial axis for both mm-scale GI-XRD and micron-scale 4D-STEM orientation maps. In addition, a weak fraction of domains oriented around particular directions are also detected by both techniques. Structural analysis were done only on low-angle misoriented areas to extract general information on epitaxial domains.

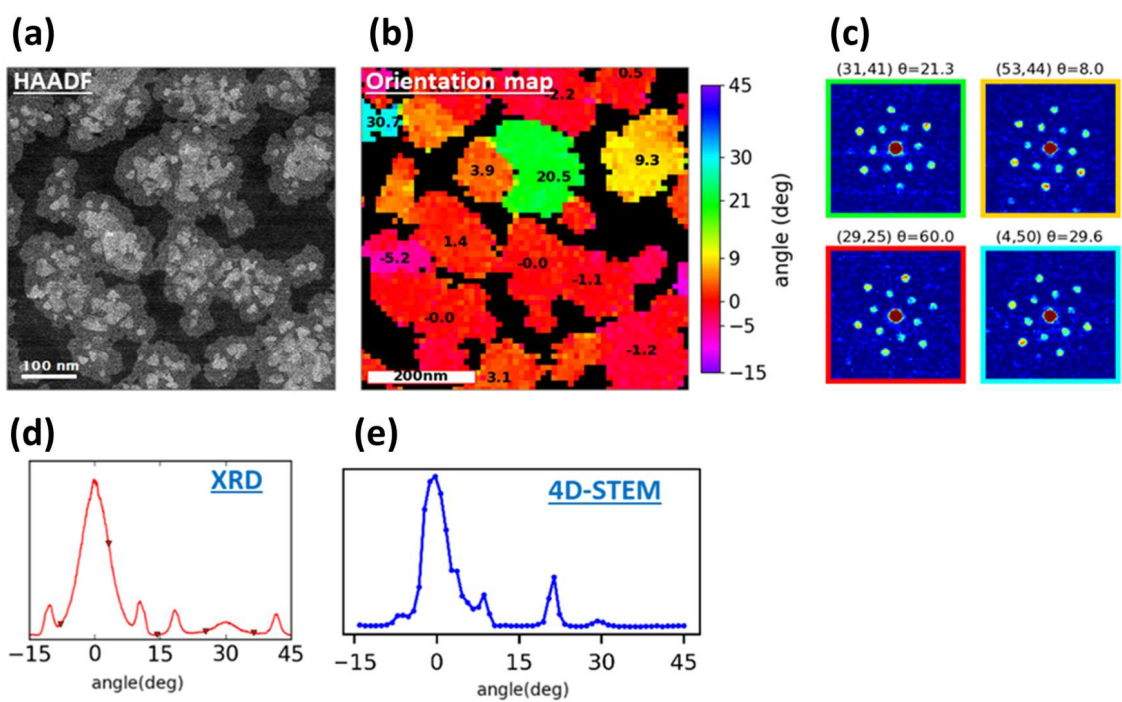

*Figure S4: WSe$_2$ grown by MBE on mica substrate and transferred on graphene support membrane (a) HAADF image, (b) orientation map, (c) examples of diffraction patterns, (d) angular distribution extracted from GI-XRD azimuthal scan and (e) 4D-STEM angular distribution of b).*

5. **Inversion domain boundaries in WS$_2$:**

The position of IDBs can be identified from 3-fold orientation maps. Figure S5 shows the linear IDBs extracted from figure 5(d) and superposed on the 6-fold orientation map shown in figure 5(g).

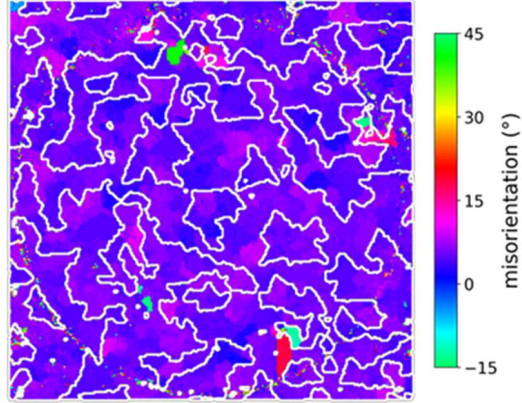

*Figure S5: Superposition of IDBs and 6-fold orientation map (from the data set of sample B shown in figure 5).*

6. **Rapid orientation mapping filter:**

For the purpose of fast orientation mapping, a custom-made Python script was developed specially for the analysis of 2D TMDs exhibiting a hexagonal diffraction pattern. The script takes as input the following quantities:

• $D_{ijhk}$ - 4D dataset, where i, j and h, k are real and reciprocal space coordinates

• $r_0$ - radial distance of the first order diffraction peaks from the center of diffraction pattern in pixels

• w - width of the diffracted spots in pixels

• s - this value is related to the slope of Fermi function used for the representation of the mask; small s corresponds to steep slope

Given the above input quantities, the script firstly creates a mask $M_{hk}$ in reciprocal space. The mask focuses on the regions of diffraction pattern where the 1$^{st}$ and 2$^{nd}$ order diffraction peaks appear. Each pixel $h, k$ is firstly assigned a complex number $z = h + ik = |z|e^{i\phi}$, where (h,k) = (0,0) corresponds to the centre of the direct beam. Furthermore, in order to reduce the impact of noise as well as the direct beam that do not carry information about orientation, a radial part of the mask is constructed as a filter that passes only the intensity values around the expected peak positions:

$$rad_1 = \frac{1}{1 + \exp(\frac{|z| - (r_0 + w/2)}{s})} - \frac{1}{1 + \exp(\frac{|z| - (r_0 - w/2)}{s})}$$

Taking into account that the 2$^{nd}$ order peaks in hexagonal 2D TMDs are found at the distance $\sqrt{3}r_0$, the second radial function is constructed as:

$$rad_2 = \frac{1}{1 + \exp(\frac{|z| - (\sqrt{3}r_0 + w/2)}{s})} - \frac{1}{1 + \exp(\frac{|z| - (\sqrt{3}r_0 - w/2)}{s})}$$

The angular part of the mask is constructed to account for the 6-fold rotational symmetry of diffraction patterns, where the angle $\phi$ represents the angle of the given pixel with respect to $h$ axis with 360° periodicity. As the second order peaks are naturally rotated by 30° w.r.t to the 1st order peaks, this shift is equally accounted for by placing a minus sign in rad$_2$:

$$M_{hk} = M_z = (rad_1 - rad_2)\left(\frac{z}{|z|}\right)^6$$

A matrix of complex numbers containing intensity and orientation information is then constructed as:

$$A_{ij} = \sum_{hk} D_{ijhk} M_{hk}$$

Finally, the image of angle of orientation θ is given by:

$$\theta_{ij} = \frac{1}{6}\arg(A_{ij})$$

The resulting image is then shown as a colour map of the orientation angle with the colour wheel spread on the range [0°, 60°]. Additionally, the intensity of diffraction at the given spot can be represented from the modulus $|A_{ij}|$ and shown as colour intensity allowing for the visualization of the number of present layers (figure S6 c)). Furthermore, a histogram of orientations is shown for a better visualization on statistical distribution of the orientation and it is directly comparable to the azimuthal scan of GIXRD. This implementation is very efficient and allows for a quick analysis, the time of analysis being ≈ 30s on standard computer for a typical 4D dataset of ≈ 1GB.

The above explained filtering can be easily adapted for the diffraction patterns with 3-fold symmetry for the purpose of detection of the local polar direction. In that case the mask is constructed as

$$M_{hk} = M_z = rad_1 \left(\frac{z}{|z|}\right)^3$$

The 2nd order diffraction peaks are not used here as they do not exhibit the 3-fold symmetry[15]. The orientation image is thus given by:

$$\theta_{ij} = \frac{1}{3}\arg(A_{ij})$$

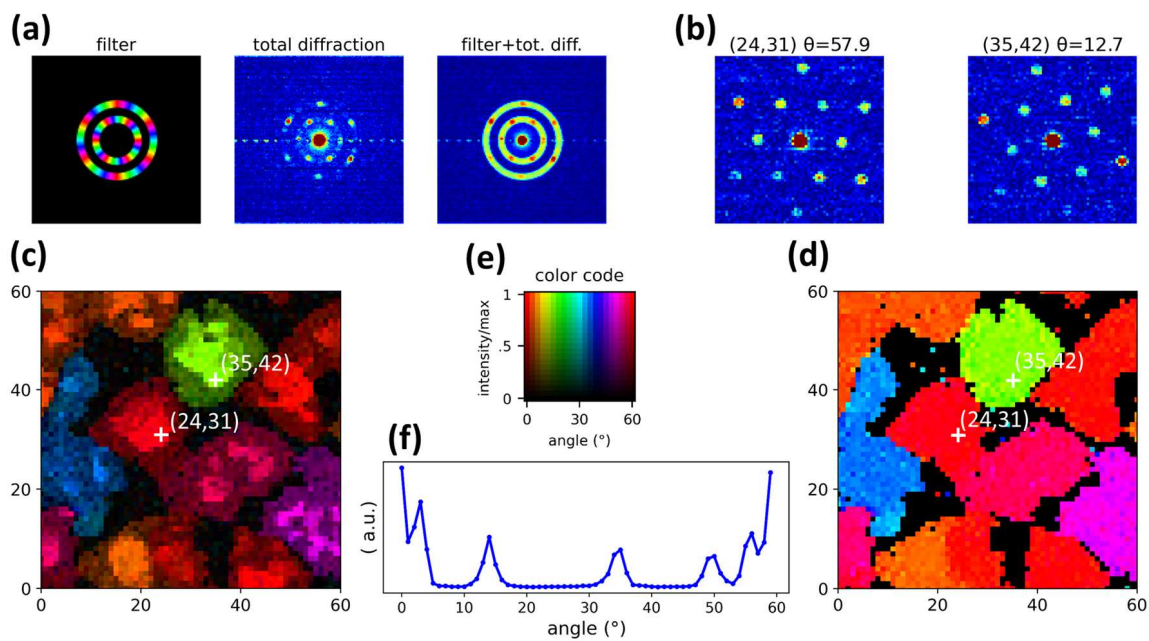

*Figure S6: (a) Filter covering 1st and 2nd order diffraction spots. (b) Example diffraction patterns and the associated orientation angles. (c): Reconstructed orientation map where the orientation angle is shown as hue and the diffraction intensity as value in the hsv color code shown in (e). (d) orientation map from (c) without distinction in diffraction intensity. (f) histogram of orientations*

### 7. Structure of WSe2: 1T':

Selenium atoms in the 1T' structure can be visualized only after Fourier filtering. The filter used here is an array type of filter that covers all the reciprocal spots of 1T' structure. The resulting image gives average atomic positions from the filtered region but the information about local point defects is lost.

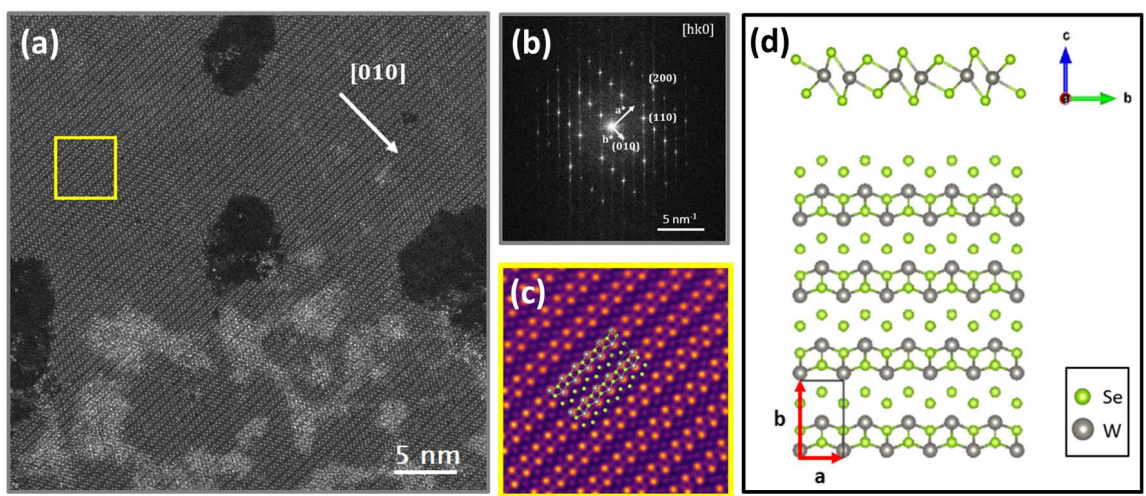

*Figure S7: (a) raw 200 kV HAADF image of $WSe_2$:1T'; (b) FFT of image (a) with indicated reciprocal lattice vectors a* and b*, (c) Fourier filtered inset from (a) showing all atomic positions including Se atoms with the overlaid model from (d), (d) model of $WSe_2$:1T' projected to (100) and (001) directions. Model visualized using VESTA[14].*

## 8. Orientation mapping of mixed phase WSe$_2$:1H-1T':

As mentioned earlier, each diffraction pattern obtained for each probe position is compared to a set of DPs corresponding to the 1H phase and having orientations in the range of [0°,60°) and to a set of template DPs corresponding to the 1T' phase and having orientations in the range [0°,180°). Due to the small size of inversion domains in WSe$_2$:1H, the electron beam is often interacting with several domains of different polarity and the anomalous contrast in the collected diffraction patterns is practically lost. For this reason, the templates corresponding to the 1H phase are constructed in the range [0°,60°). From the point of view of the phase transformation 1H→1T', there are three equivalent orientations of WSe$_2$:1T'. The same amount of atomic movements is required for a phase transformation from a single 1H region to any of the three orientational variants of 1T' (figure S9). The orientation mapping from template matching is here represented with the color code corresponding to the range [-90°, 90°). The orientation of WSe$_2$:1H phase takes exclusively value from the range [-30°, 30°). The actual orientation distribution is even narrower, centered around the principal peak and exhibits two characteristic peaks at about -11° and +11°. This region is thus represented in shades of red color. On the other hand, the WSe$_2$: 1T' region is represented as shades of 3 principal colors: red, blue and green corresponding to misorientations of 0°, -60° and +60°. As seen from the histogram of misorientations, the three principal orientations are almost equally distributed in the scanned region and exhibit a shape similar to the one observed for WSe$_2$:1H orientational distribution.

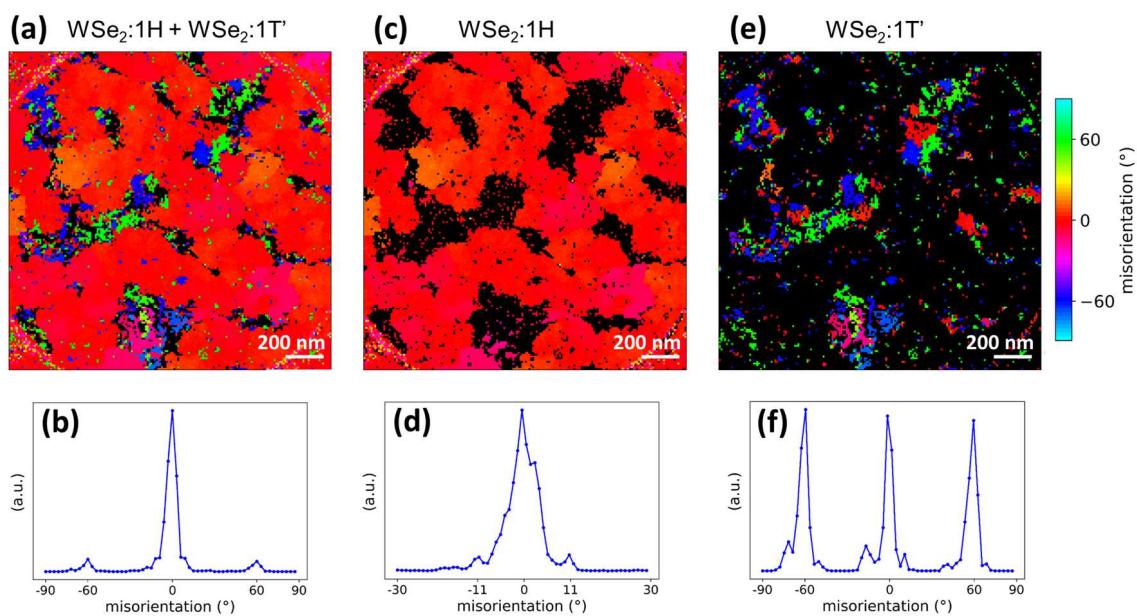

*Figure S8: (a) The orientation map of heterophase WSe$_2$:1H-1T' sample done by template matching algorithm represented in the range [-90°, 90°) and (b) the corresponding histogram of orientations. (c) The orientation map of WSe$_2$:1H phase and (d) the corresponding histogram of orientations. (e) The orientation map of WSe$_2$:1T' phase and (f) the corresponding histogram of orientations.*

As the three principal misorientations are equal in the sense of 1H→1T' phase transformation, to prove the hypothesis of phase transformation against the independent nucleation, the misorientation θ of WSe$_2$:1T' is shown as θ mod 60. This can be done either mathematically from the results of template matching or by running a rapid orientation mapping algorithm while choosing a filter around a hexagonal reciprocal sublattice of 1T'. The comparison of the two approaches is given in figure S10. The two approaches give comparable orientation maps and corresponding histograms. However the template matching procedure here is much more sensitive to any noise in diffraction pattern. This can be seen at the edges of circular hole of the TEM grid, where the orientation is not determined correctly due to the additional signal appearing from the scattering of the edge of the TEM grid.

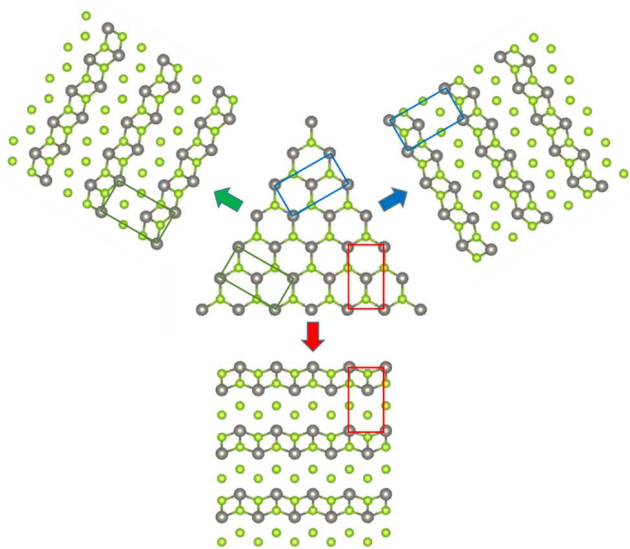

Figure S9: Three grains of WSe$_2$:1T' with different orientations formed from a single grain of WSe$_2$:1H. Model visualized using VESTA[14].

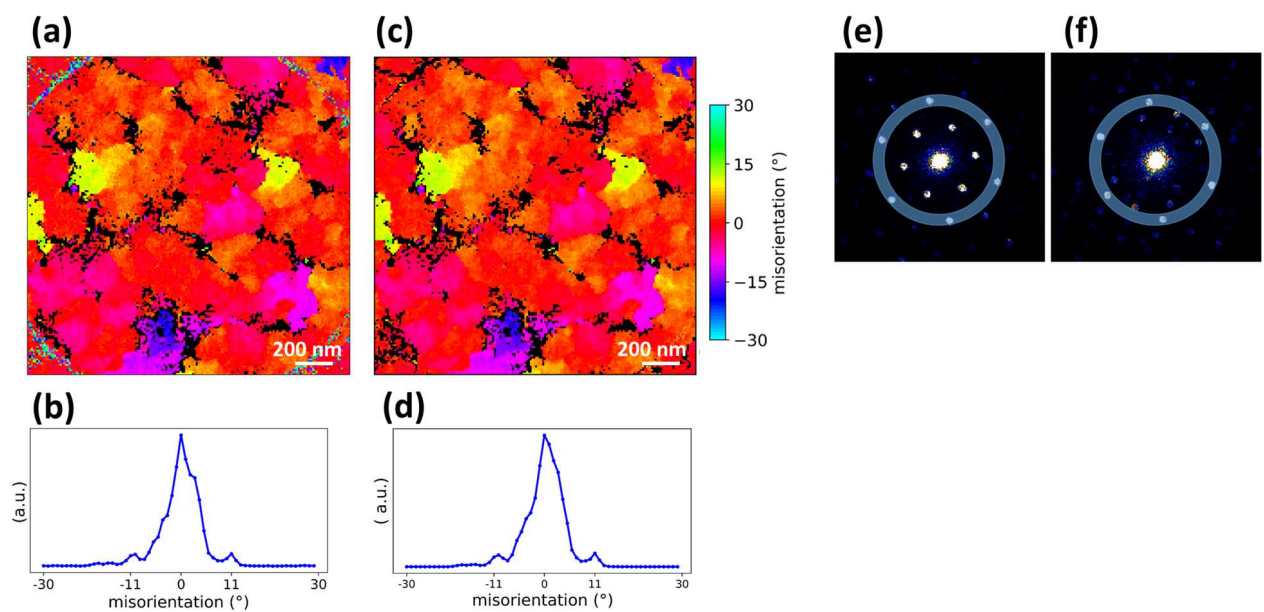

Figure S10: (a-b) Modulo 60 (θ mod 60) of the template matching based orientation map shown in figure S8(a) and the corresponding histogram of orientations. (c-d) orientation map and the corresponding histogram of orientations based on the rapid orientation mapping using the radial filter covering hexagonal spots common to the two phases indicated on the example diffraction patterns of (e) 1H phase and (f) 1T' phase.